# Interface Weyl electrons in the kagome-bilayers

K. Ishida[1,*], K. Fujiwara[2], S. Sasano[3], K. Nakazawa[4,5], T. Seki[6,7,8], S. Souma[7,9], D. Shiga[10,11], H. Kumigashira[10,11], T. Sato[7,8,9,12], R. Ishikawa[3], N. Shibata[3,13], Y. Motome[1], A. Tsukazaki[1,4,6,14,*]

[1] Department of Applied Physics, The University of Tokyo, Bunkyo, Tokyo, Japan.
[2] Department of Chemistry, Rikkyo University, Tokyo, Japan.
[3] Institute of Engineering Innovation, The University of Tokyo, Bunkyo, Tokyo, Japan.
[4] RIKEN Center for Emergent Matter Science (CEMS), Wako, Japan.
[5] RIKEN Center for Quantum Computing (RQC), Wako, Japan.
[6] Institute for Materials Research (IMR), Tohoku University, Sendai, Japan.
[7] Center for Science and Innovation in Spintronics (CSIS), Tohoku University, Sendai, Japan.
[8] International Center for Synchrotron Radiation Innovation Smart (SRIS), Tohoku University, Sendai, Japan.
[9] Advanced Institute for Materials Research (WPI-AIMR), Tohoku University, Sendai, Japan.
[10] Institute of Multidisciplinary Research for Advanced Materials (IMRAM), Tohoku University, Sendai, Japan.
[11] Photon Factory, Institute of Materials Structure Science, High Energy Accelerator Research Organization (KEK), Tsukuba, Japan.
[12] Department of Physics, Graduate School of Science, Tohoku University, Sendai, Japan.
[13] Nanostructures Research Laboratory, Japan Fine Ceramics Center, Nagoya, Aichi, Japan.
[14] Quantum-Phase Electronics Center (QPEC), The University of Tokyo, Tokyo, Japan.

*e-mail : K. I.: hitagi-keiki443@g.ecc.u-tokyo.ac.jp & A.T.: tsukazaki@ap.t.u-tokyo.ac.jp,

**Abstract**

What makes a topological character prominent is whether its characteristic electronic states are located near the Fermi energy. While the kagome-metals possess unique topological characteristics owing to their distinctive electronic band structures, i.e. flat and Dirac bands, the control of Fermi energy to accentuate these characteristics has scarcely been studied in these materials. Here, we propose a new concept for activating the topological characteristics of Weyl electrons specifically near the interface of kagome-bilayers, namely interface Weyl electrons. The kagome-bilayers prepared by a sputtering technique surprisingly exhibit atomically abrupt interfaces thanks to the suppression of interdiffusion for maintaining the crystal symmetry of the hexagonal-type $D0_{19}$ and cubic-type $L1_2$ structures. Much higher electrical conductivity of the kagome-bilayers compared to single-layer films indicates the emergent contribution of a highly conductive electronic state at the interfaces. The enhanced electrical conductivity and the exceptionally large anomalous Hall conductivity about 2000 $\Omega^{-1}cm^{-1}$ in the intrinsic mechanism indicates that the emergent interface electronic states comprise Weyl nodes in the $D0_{19}$ structure identified in our theoretical band calculations, triggered by nontrivial charge transfer. These results demonstrate that topological transport can be engineered at the interface of intermetallic compounds, leading further design principles towards development of device functionality.

## Main

Interface engineering is a vital technique for developing the functionality of a wide range of devices using different materials, for example semiconductor two-dimensional systems[1], magnetoresistance devices[2,3], topological insulator heterostructures[4,5], and recent van der Waals (vdW) stacking[6-9]. While electronic interaction plays various roles for each functionality via charge transfer, magnetic coupling, and proximity effect, it remains difficult to theoretically predict modifications in electronic states that are suitable for highlighting topological properties at interfaces. As an exception, graphene-stacked heterostructures where two types of close-packed stacking, ABAB (Bernal) and ABC (rhombohedral), are freely controlled, leading to distinct band structures[10,11], superconductivity[12-14], and topological electronic states[15-18]. In the topological materials, while significant progress has been made in classifying topology in the band structure of materials through theoretical studies incorporating crystal symmetry and spin-orbit interactions[19-21], the advances in experimental techniques for controlling topology and Fermi energy have been limited to specific materials. The discovery of topological material families, such as topological insulators and Dirac and Weyl semimetals, presents a further challenge in manifesting distinct phenomena through the Fermi energy control. The control schemes for the electronic state, as like stacking control in graphene, implemented by interface engineering will significantly contribute to the activation of topological characteristics and lead to the development of interface device functions that utilize their unique features, such as in intermetallic kagome metals.

Kagome metals constitute a class of topological materials exhibiting a variety of remarkable properties in terms of magnetic, superconducting, and topological phenomena[22-35], which originate from their characteristic geometries of two-dimensional kagome lattices composed of 3*d* transition metals (Fig. 1a). Unlike van der Waals materials, however, the kagome layer cannot be exfoliated owing to its strong interlayer bonding, even though such two-dimensional kagome lattices form layered stacking sequences in intermetallic alloys. The kagome lattice hosts the distinguished band structures with flat bands, van Hove singularities, and Dirac or Weyl bands with large Berry curvature (Fig. 1b), stimulating intense interest in both fundamental physics and device applications[22,23,35], yet most investigations so far have focused on bulk crystals and single-phase single-layer films. An important challenge lies in exploiting the control of the electronic states and/or the Fermi energy to manifest the topological characteristics driven by the Berry curvature and the large density of states arising from flat bands or van Hove singularities. By applying conventional chemical substitution technique (for example Pt substitution into $Co_3Mo$ as $Co_3Mo_{1-x}Pt_x$ alloy film in Fig. 1c), it is not straightforward to vary the Fermi energy because the structural stacking degree of freedom allows segregated layers. In fact, as shown in Figs. 1c and 1d, kagome-layer stacking in the closest-packed structure has two degrees of freedom, which naturally leads to the formation of various layered polymorphs such as ABC, AB, and BC layered stacking. Conversely, the artificial kagome-layer-stacked interface, which overcomes the difficulty of the non-delaminable feature, is an unexplored approach to manifesting the topological phenomena based on a new degree of freedom with stacking control and/or as an alternative to conventional chemical substitution.

Here we exemplify the concept of emergent interface electrons at the atomically abrupt stackings in kagome-bilayers. The synthesized bilayers by sputtering exhibit significant electrical conductivities far exceeding other kagome-metals and a large anomalous Hall conductivity arising from an intrinsic origin. These results, which are not pronounced in the single-layer films, indicate that the large Berry curvature associated with the Weyl nodes manifest specifically near interface with finite width induced by the band bending via nontrivial charge transfer. These findings in the kagome-bilayers demonstrate that various intermetallic alloys with topological electronic states can serve as an intriguing platform for exhibiting remarkable modification exclusively at the interface.

## Design of kagome-bilayers

The concept of this study is the extension of stacking engineering to the intermetallic kagome-metals. The platform of the kagome-lattice-layer is based on $Co_3X$ ($X$ = Mo, Pt), showing the fundamental lattice structure consisting of cobalt atoms in kagome lattice with $X$ atoms occupying the centers of the hexagons (Fig. 1a). There are two types of stacking sequences of kagome layers, that is ABC (Fig. 1e) and ABAB (Fig. 1f) corresponding to $L1_2$ and $D0_{19}$ structure, respectively. Although these two structures are composed of the same kagome-layer building blocks, they are classified into distinct symmetries of hexagonal and cubic. One example of $D0_{19}$ is $Co_3Mo$, exhibiting room-temperature ferrimagnetism with perpendicular magnetic anisotropy along [0001] direction[36]. The other example is $Co_3Pt$ for $L1_2$ structure, which is a room-temperature ferromagnet[37]. Notably, this symmetrical difference due to stacking sequences leads to markedly different electronic band structures, i.e. the topological flat and Dirac bands are only apparent in $D0_{19}$ structure (Extended Data Fig. 1). This distinction may be reminiscent of the recent studies on ABAB Bernal and ABC rhombohedral multilayer graphene, in which different stacking configurations generate fundamentally distinct band structures in terms of topology[9,10]. Although the crystal structure is distinct, it will be interesting to see how the interface forms because these crystallographic polymorphs share a common cobalt-based kagome framework. We designed a heterostructure consisting of kagome-layer stacking in the [0001] direction of $D0_{19}$ structure and the [111] of $L1_2$ (Fig. 1g).

The single-layer films and bilayers were synthesized on $Al_2O_3$ substrates by sputtering method[36] (Methods). To make this concept experimentally, we exploited the alloy films of $Co_3Mo_{1-x}Pt_x$, whose stable structure can be controlled by Mo/Pt composition ratio as shown in Extended Data Fig. 2 (ref. 36). As increasing the Pt content $x$, the stable structure transits from the $D0_{19}$ to the $L1_2$ phase (Extended Data Fig. 2), which was examined by x-ray diffraction. Clear Laue fringes indicate the flat surface and highly crystalline uniformity with minimal defects (Supplementary Fig. 1c). With applying the composition-controlled and in-plane-lattice matched $Co_3Mo_{1-x}Pt_x$ films (Extended Data Fig. 3), the kagome-layer was intentionally stacked for making heterointerfaces.

**Atomically abrupt kagome-layer stacking**

The kagome-layer stacking (Fig. 2a) of $L1_2$ ($x$ = 0.44) / $D0_{19}$ ($x$ = 0.08) was observed by scanning transmission electron microscopy (STEM) to investigate the interface stacking. Figure 2b presents an atomic-resolution annular dark field (ADF) STEM image. The interface is atomically abrupt in a large field-of-view image (Extended Data Fig. 4a), enabling us to distinguish the sequence of kagome-layer stacking in the vicinity of the interface. The lower region exhibits the AB stacking sequence of the $D0_{19}$ structure, whereas the upper region displays the ABC stacking sequence of the $L1_2$ structure (marked by red lines in Fig. 2b). At the interface, however, the location of the boundary layer, corresponding to the interface between $D0_{19}$ and $L1_2$ is not obvious. Since the crystal structures of $D0_{19}$ and $L1_2$ are defined by stacking sequences of three or more kagome-layers, the definition of the interface is non-trivial (see white broken lines and arrows upward AB- and downward ABC-stacking). Judging from the atomic-resolution energy dispersive x-ray spectroscopy (EDS) maps of Co (green in Fig. 2c) and Mo (red) and Pt (pink) (Fig. 2d), the boundary layer can be identified as "C-layer" owing to the apparent transition from Mo-rich at lower side to Pt-rich at upper side. The suppression of Pt interdiffusion, even in compositions that are similar in both layers, is likely due to the large phase stability of both structures, in other words, structural symmetry protection. The presence of an atomically abrupt kagome-metal interface with suppressed interdiffusion demonstrates the thermodynamic stability of the kagome-layer-stacked $L1_2$/$D0_{19}$ interface, despite the high-temperature processes of synthesis and annealing.

We further performed the electron energy-loss spectroscopy (EELS) analysis of Co-$L_{2,3}$ edge at seven layers across the interface (see also fifteen layers data in Extended Data Fig. 4b). The spectral feature remains largely preserved (Fig. 2e), indicating that the electronic state of Co evolves continuously across the interface. The peak positions of the Co-$L_3$ edge are plotted in Fig. 2f, clearly revealing a systematic chemical shift by approximately 0.3 eV. The observed shift is substantially larger than the expected from local distortions in coordination environment and therefore points to a significant modification of the Co 3*d* electronic states. The Co-$L_3$ edge of the $L1_2$ layer is located at lower energy loss than that of the $D0_{19}$ layer, indicating a higher occupation of Co 3*d* states in the $L1_2$ phase. This energy shift mainly comes from the differences of the Fermi energy position and the occupied electronic bands owing to different crystal structure (Extended Data Fig. 1). Although direct band-resolved evidence is limited by an inevitable alloy-disorder-induced broadening for ARPES in the substituted films with less defects (Supplementary Fig. 2), these EELS observations provide direct evidence of the electron charge transfer from Co atoms in the $L1_2$ layer to those in the $D0_{19}$ layer, possibly leading to substantial charge accumulation into the higher energy bands above $E_F$ of $D0_{19}$ layer.

## Emergent high interface conduction

The kagome-bilayers of both the $L1_2/D0_{19}$ and $D0_{19}/L1_2$ exhibit an unexpected enhancement of electrical conduction compared to the single-layer films of $D0_{19}$ ($x = 0$) and $L1_2$ ($x = 0.48$). Figure 3a presents the temperature dependence of the sheet conductivity $\sigma_{\text{sheet}}$ of those samples having comparable total thickness about 40 nm. In addition to good metallic behaviour of all samples, both bilayers furthermore exhibit a substantially large sheet conductivity than both single-layer films, indicating the emergence of an additional conduction channel at both stackings of $L1_2/D0_{19}$ and $D0_{19}/L1_2$ in the bilayer. For comparison with the values of other kagome-metals, the thickness-averaged resistivity $\rho_{xx}^{\text{ave}}$ is plotted as a function of temperature in Fig. 3b. The value of room-temperature $\rho_{xx}^{\text{ave}}$ for the bilayer is the lowest one of those reported for single crystals of several representative kagome-metals especially $D0_{19}$ structure and cobalt-kagome magnets[24,26,34,38,39], highlighting the exceptionally high room-temperature conductivity of the interface electrons. In addition, thickness-controlled bilayers also exhibit constantly low $\rho_{xx}^{\text{ave}}$ in Fig. 3c compared to the increase trend of single-layer films in thickness reduction (see also Extended Data Fig. 5). To investigate the effective thickness of the interface conductive channel, we calculate the three-dimensional resistivity of the interface $\rho_{xx}^{\mathrm{int}}$ with the assumed thickness of the conducting channel $t_{\mathrm{cal}}^{\mathrm{int}}$ in Fig. 3d. The sheet conductance of the interface $\sigma_{\mathrm{sheet}}^{\mathrm{int}}$ ($\Omega^{-1}$) is estimated by the difference of sheet conductance of the bilayers $\sigma_{\mathrm{sheet_bilayer}}$ (average of $L1_2$/$D0_{19}$ and $D0_{19}$/$L1_2$) and that of single-layer films $\sigma_{\mathrm{sheet_bulk}}$ (average of $D0_{19}$ and $L1_2$) shown in Fig. 3a. Following the simple equation $\rho_{xx}^{\mathrm{int}}$ ($\Omega$cm) = $t_{\mathrm{cal}}^{\mathrm{int}}$ / $\sigma_{\mathrm{sheet}}^{\mathrm{int}}$, this value suggests the quantitative consideration for the thickness of the interface conducting channel. As shown in Fig. 3d, when the conducting region is assumed to spread to 3 nm, 4 nm and 12 nm, the $\rho_{xx}^{\mathrm{int}}$ value is comparable to that of Cu, Al and Ru, respectively. These assumed thicknesses are rather thick under consideration based on the screening length of the conventional metals with $10^{22}$ cm$^{-3}$ electrons, approximately 0.1 nm. In another experimental results about charge transfer in Fig. 2f, the slope of Co $L_3$ edge shows transition in four kagome-layers corresponding to roughly 1 nm, which is also very thin. Although it is challenging to experimentally support this, it is possible that this conductive layer is thin, resulting in extremely low $\rho_{xx}^{\mathrm{int}}$. On the other hand, it may also be important to consider the possibility of a significant expansion of the conducting region, given that charge transfer may occur preferentially in specific semimetallic orbitals. In any case, it is extremely difficult to evaluate the thickness of the conducting region experimentally in such metal-bilayers, so at this stage, it remains a speculation of nontrivial charge transfer. Since high electrical conductivity has been experimentally observed in the kagome-bilayers, they could be used to improve future devices such as interconnect conductors.

**Large anomalous Hall conductivity**

Hall effect and magnetization measurements were further carried out. In top and bottom panels of Figs. 4a–c, the magnetic field $\mu_0 H$, where $\mu_0$ is vacuum permeability, dependence of magnetization $M$ and the thickness-averaged Hall conductivity $\sigma_{xy}^{\mathrm{ave}}$ at $T$ = 300 K is compared for three samples, respectively. Magnetization behaviour is consistent with the previous studies[36], indicating that single-layer films of $D0_{19}$ and $L1_2$ are easy to be magnetized in the out-of-plane (Fig. 4b) and in-plane (Fig. 4c) directions, respectively. For the magnetization of bilayer, the shape of *M-H* curve can be understood as sum of two consisting layers, out-of-plane magnetized bottom $D0_{19}$ and in-plane magnetized top $L1_2$ layer, which was examined by magneto-optical Kerr effect measurements at front and back sides (Extended Data Fig. 6). The coercive field in the hysteresis of the $\sigma_{xy}^{ave}$ well agrees with that of magnetization. In clear contrast to the comparable saturation values of magnetization in these samples, the $\sigma_{xy}^{ave}$ of the bilayer presents clear enhancement from that of both single layers. Since $R_{yx}$ saturates above critical fields as consistent as magnetization, the ordinary Hall component is negligibly small owing to metallic Fermi surfaces (Supplementary Fig. 3). The $\sigma_{xy}^{ave}$ of the $L1_2/D0_{19}$ bilayer is distinctly large as 1800 $\Omega^{-1}cm^{-1}$ (see also the results of $D0_{19}/L1_2$ in Extended Data Fig. 7). Judging from the hysteresis shape of $\sigma_{xy}^{ave}$, the charge carrier in the $D0_{19}$ layer forming hysteresis contributes dominantly large value about 1500 $\Omega^{-1}cm^{-1}$ of the emergent anomalous Hall conductivity to total 1800 $\Omega^{-1}cm^{-1}$. Moreover, the large $\sigma_{xy}^{ave}$ of both bilayers persists over a wide temperature range from 300 to 2 K owing to high magnetic transition temperature[36] as shown in Fig. 4d, in which region the $\rho_{xx}^{ave}$ shows good metallic behaviour in Fig. 3b. Following this trend, the scaling analysis in Fig. 4e summarizing all temperature data highlights that the $\sigma_{xy}^{ave}$ are sufficiently high at constant along $\sigma_{xx}$ variation in the region governed by intrinsic mechanism[40], comparable to that of representative magnetic metals and Weyl magnets $Co_3Sn_2S_2$ (refs.[26,41]). These results provide reliable evidence for an intrinsic Berry-curvature origin of the anomalous Hall effect in the bilayer. In contrast, the small values of $\sigma_{xy}^{ave}$ for the single-layer films are not significant, although the value is consistent to the theoretical one approximately 80 $\Omega^{-1}cm^{-1}$ at $E = 0$ eV in right panel of Fig. 4f. First-principles calculations of $D0_{19}$-type $Co_3Mo$ with the ferrimagnetic order reveal that the positive large value of 1000 $\Omega^{-1}cm^{-1}$ appears at higher energy around 0.3 eV above the $E_F$ (red shaded region in Fig. 4f). The large values of $\sigma_{xy}^{ave}$ likely originate from the Berry-curvature contributions at the gapped topological states under the broken time reversal symmetry[40,42], for example at high symmetric K point and around the K-Γ line (Fig. 4g). Considering the symmetry of band structure and spin-orbit coupling (Extended Data Fig. 8), the gapped feature along K-Γ line disappears along Γ-M line in Fig. 4g, indicating the formation of Weyl nodes driven by the band crossing as similar as the evidence for Weyl character in $Co_3Sn_2S_2$ (ref.[36]) and $Mn_3Sn$ (ref.[43]). We discovered that the consistently large anomalous Hall conductivity of the specific Weyl nodes is pronounced around the interface of kagome-bilayers even no feature in single-layer films.

## Discussion

The apparent enhancement of electrical sheet conductance (Fig. 3) and anomalous Hall conductivity (Fig. 4) in the bilayer provides compelling evidence that a highly conductive electrons emerges at the kagome-metal interface. Here we discuss the topological characteristics of the interface electrons. First, the dramatic enhancement of $\sigma_{xy}$ in the bilayer can plausibly be attributed to the contribution from the source of Berry curvature in $D0_{19}$-type $Co_3Mo$ under broken time-reversal symmetry (Fig. 4g), i.e. massive Dirac state[25] or Weyl nodes[40,42], identified in our band calculations. These electronic states are activated by band bending via charge transfer as discussed in Fig. 2f regarding the Co-$L_3$ peak shifts of about 0.3 eV. Considering the calculated bands exhibiting large $\sigma_{xy}$ around 0.25~0.38 eV recalculated to 0.2~0.29 eV via mass-renormalization factor of 1.3 (ref. 36), the quantitative agreement in energy scale of the band bending supports this scenario. In addition, the large density of states at the identical energy position (Extended Data Fig. 1) also supports the enhancement of the $\sigma_{sheet}$. Second, although the energy shift can be verified as discussed above, the effective thickness of the charge transfer, related to the dimensionality of the interface electrons, is nontrivial at metallic bilayer, as discussed in Fig. 3d. While the Weyl states are generally defined in three-dimensional bands, the anisotropic three-dimensional bands in the kagome-lattice may host Weyl electrons within an interfacial region of finite width, as similar to the critically thin criterion exhibiting Weyl semimetal of layered $WTe_2$ (ref.[44]) and two-dimensional Weyl semimetal Bismuthene/SnS (ref.[45]). Direct experimental verification of the Weyl nodes and associated surface Fermi arcs remains challenging as the relevant electronic states are buried near the interface. Nevertheless, our theoretical calculations for $Co_3Mo$ and the observed negative magnetoresistance (Extended Data Fig. 9) strongly support the Berry curvature contribution to the experimentally observed transport results. The critical parameter for hosting Weyl feature in low dimension needs to be addressed by theoretical and experimental future approaches. We therefore conclude that the Weyl electrons are activated by charge transfer specifically near the interface, namely interface Weyl electrons, which exhibits the intriguing topological transport properties. The concept of topological properties only manifesting at interfaces will lead to the discovery of characteristics that remain hidden in single layers of many material groups.

## Methods

**Sample preparation:** $Co_3Mo_{1-x}Pt_x$ thin films and kagome-bilayers of $L1_2$ (20-nm)/$D0_{19}$ (20-nm) and $D0_{19}$ (20-nm)/$L1_2$ (20-nm) were fabricated on $Al_2O_3$ (0001) substrates by radio-frequency magnetron sputtering method[36]. $Al_2O_3$ substrates were annealed at $T$ = 1000 °C for 2 hours to obtain step and terrace structure. After the films were deposited at $T$ = 500 °C, the films were annealed at $T$ = 800 °C for 3 hours. Thicknesses of the films were controlled by duration, which was evaluated by x-ray Laue fringes. The films were capped with $SiO_x$ (approximately 50 nm) to suppress oxidation in air except for ARPES measurement. The composition of the films was regulated by target compositions with Mo and Pt chips on Co plate. The lattice parameters were evaluated by X-ray diffraction with Cu Kα1 (wavelength : 1.54056 Å) with Empyrean (Malvern Panalytical). Composition of the films were characterized by energy-dispersive X-ray spectroscopy (JEOL).

**Band structure and anomalous Hall conductivity calculations:** Electronic structure calculations were performed within the framework of density functional theory using the OpenMX code. The exchange–correlation functional was treated within the generalized gradient approximation in the Perdew–Burke–Ernzerhof form. Norm-conserving pseudopotentials were employed, and the Kohn–Sham wave functions were expressed as linear combinations of pseudoatomic orbitals, with the basis sets Co6.0H-s3p2d1, Mo7.0-s3p2d2, and Pt7.0-s3p2d2f1 selected for Co, Mo, and Pt atoms, respectively. Fully relativistic calculations including spin–orbit interaction[46] were conducted to obtain the electronic structures. In-plane and out-of-plane lattice parameters were fixed to 5.145 Å and 4.121 Å for $Co_3Mo$, and 5.193 Å and 4.240 Å for $Co_3Pt$, based on experimental values. A cutoff energy of 1200 Ry was used for the Fourier transform grid, and the Brillouin zone was sampled using a k-point mesh of 31 × 31 × 31 for self-consistent calculations. The anomalous Hall conductivity was calculated using OMXsigmaxy[47], a post-processing code for OpenMX, based on the expanded Fukui–Hatsugai–Suzuki method[48], with a k-point mesh of 256 × 256 × 256.

In supplementary Fig. 4, the heterostructure of ABCAB, corresponding a model of $L1_2/D0_{19}$ stacked interface, is presented.

**Scanning transmission electron microscopy:**

The electron-transparent thin specimen for STEM observation was prepared using a focused ion beam system (FIB; Helios 5, Thermo Fisher Scientific, USA). A small portion of the sample was cut and lifted onto a copper grid, followed by thinning with a Ga ion beam. To remove surface damage layers, Ar ion milling at 0.5 kV was performed using NanoMill (Fischione Instruments) in the final stage. ADF STEM images were acquired by ARM200F (JEOL, Ltd.) equipped with an ASCOR corrector and a cold field-emission gun, operated at 200 kV. The convergence and collection semi-angles were 24 mrad and 53-175 mrad, respectively. EEL spectra were acquired by a Continuum spectrometer (Gatan, Inc.), equipped with ARM 200F microscope (JEOL Ltd.). The convergence and collection semi-angles were 24 and 62 mrad, respectively. To enhance the signal-to-noise ratio, the sequential averaging method was used for ADF-STEM images[49], and total third-degree variation (TTDV) regularization was used for EDS elemental maps[50].

**Electrical transport measurement:** Resistivity and Hall effect were measured by Physical Properties Measurements System (Quantum Design Inc.). The films were manually scratched to Hall-bar shape with attaching In electrical contact. $\sigma_{xx}^{ave}$ and $\sigma_{xy}^{ave}$ for the bilayers were calculated with thickness and conventional equations $\sigma_{xx}^{ave} = \frac{\rho_{xx}^{ave}}{(\rho_{xx}^{ave})^2+(\rho_{yx}^{ave})^2}$ and $\sigma_{xy}^{ave} = \frac{\rho_{yx}^{ave}}{(\rho_{xx}^{ave})^2+(\rho_{yx}^{ave})^2}$. To estimate the value of anomalous Hall conductivity at saturated magnetization condition, we did not subtract the ordinary Hall component due to negligibly small contribution (Supplementary Fig. 3). Temperature dependence of anomalous Hall angle, $\sigma_{xy}^{ave}/\sigma_{xx}^{ave}$, is compared with that of $\sigma_{xx}^{ave}$ and $\sigma_{xy}^{ave}$ for two bilayers of $L1_2/D0_{19}$ and $D0_{19}/L1_2$ (Supplementary Fig. 5).

**Magnetic properties measurements:** Magnetic field dependence of magnetization was measured at $T$ = 300 K by VersaLab (Quantum Design Inc.). Magneto-optical Kerr effect measurement was carried out with 680 nm laser under the out-of-magnetic field application (Neoark Corporation).

**Angle-resolved photoemission spectroscopy:** After the thin films were fabricated in a vacuum chamber using the sputtering technique, they were extracted into air and quickly placed in a vacuum-sealed tube. The films were transferred to the Photon Factory of the High Energy Accelerator Research Organization (KEK-PF) with a vacuum-sealed tube, attached to the ARPES sample holder in the glovebox, and then installed into the ARPES chamber without exposure to air. Clean surface was obtained by annealing of the film at 800 °C for 3 hours. Soft-X-ray-ARPES measurements were carried out with an Omicron-Scienta SES2002 electron analyzer with energy-tunable synchrotron light at BL-2A in KEK-PF. We used linearly polarized light (horizontal polarization) of 250-1020 eV, with an energy resolution of 100-170 meV. All the ARPES measurements were conducted at $T$ = 50 K in the ferrimagnetic phase.

**Data availability**

The data that support the findings of this study are available from the corresponding author upon request.

**Acknowledgements**

A part of this work is supported by JSPS KAKENHI (Grant No. 22H00288, 25H01250, 25H01247, 25H00598, 24K00564, 24K17518, 24H00373, 25K21694, 25H00793), JST-FOREST (Grant No. JPMJFR223Y, JPMJFR2033), JST-ERATO (Grant No. JPMJER2202), KEK-PF (Proposal number: 2024G141, 2024S2-001), "Advanced Research Infra-structure for Materials and Nanotechnology in Japan (ARIM)" of the Ministry of Education, Culture, Sports, Science and Technology (MEXT), (Grant No. JPMXP1226UT0052), and MEXT Initiative to Establish Next-generation Novel Integrated Circuits Centers (X-NICS) Grant Number JPJ011438.

**Author contributions**

A.T. conceived the project. K.I. fabricated and characterized the films with the support of K.F.. K.N. and Y.M. conducted the ab initio calculation. S.S., R.I., and N.S. carried out the STEM measurements. K.I. and K.F. carried out the magnetization measurement with the support of T. Seki. S.S., D.S., H.K, and T. Sato carried out the ARPES measurement and analysis of the data. K.I. and A.T. wrote the manuscript with input from all the authors.

**Competing interests**

The authors declare no competing interests.

## Figure captions

**Fig. 1 | Design principle for emergent interface electrons in kagome-bilayers. a,** A schematic of kagome-layer composed of cobalt Co and *X*. **b,** Simplified schematic of band structure of single-layer kagome lattice without spin-orbit coupling, exhibiting the Dirac band at the K point and a flat band across the whole Brillouin zone. Broken line indicates Fermi energy $E_F$, which depends on composition and structures. **c,** Field-of-view ADF STEM image of the mixed-alloy $Co_3Mo_{1-x}Pt_x$ film ($x$ = 0.31). Various stacking AB-, ABC-, and BC-stacking are marked. **d,** A schematic of kagome-layer stackings. **e,** ABC-stacking of the kagome-layer forming cubic $L1_2$ structure. **f,** AB-stacking of the kagome-layer forming hexagonal $D0_{19}$ structure. **g,** A schematic of kagome-bilayer structure consisting of two types of kagome metals. Interface electronic states appeared in the kagome-bilayers.

**Fig. 2 | Charge transfer at atomically abrupt interface in kagome-bilayer. a,** A schematic of bilayer consisting of top $L1_2$-layer and bottom $D0_{19}$-layer with charge transfer depicted at the interface. **b,** Atomic-resolution ADF STEM image at the $L1_2/D0_{19}$ interface, viewed along the $[11\bar{2}0]$ and $[1\bar{1}0]$ directions of the $D0_{19}$ and $L1_2$ structures, respectively. Each layer is defined as A, B, and C layers, forming AB-stacking and ABC-stacking. Colored circles correspond to Co (green), Mo (red), and Pt (Purple). **c and d,** TTDV denoised EDS elemental maps at the interface for Co (**c**) and Pt and Mo (**d**). **e,** EEL spectra of Co-$L_{2,3}$ edges at seven points across the interface. **f,** Peak positions of Co-$L_3$ edges across the interface.

**Fig. 3 | Emergent highly conductive channels at the kagome-bilayer interfaces. a,** Temperature dependence of sheet-conductivity $\sigma_{\mathrm{sheet}}$ in the $L1_2/D0_{19}$ and $D0_{19}/L1_2$ bilayers ($t_{\mathrm{total}}$ = 41 nm) compared to single-layer films of the $D0_{19}$ structure ($x$ = 0, $t_{\mathrm{total}}$ = 46 nm) and $L1_2$ structure ($x$ = 0.48, $t_{\mathrm{total}}$ = 46 nm). **b,** Corresponding thickness-averaged resistivity $\rho_{xx}^{\mathrm{ave}}$ of the bilayers and the single-layer films shown in **a**. The resistivity for the representative bulk single crystals of $Fe_3Sn_2$, $D0_{19}$-type kagome metals and cobalt-kagome metals are compared. The

values are from $Fe_3Sn_2$ (ref.[38]), $Mn_3Sn$ (ref.[24]), $Ni_3In$ (ref.[34]), $Ni_3Sn$ (ref.[34]), $Co_3Sn_2S_2$ (ref.[26]), and CoSn (ref.[39]). **c,** Total thickness dependence of the $\rho_{xx}^{ave}$ of the single-layer films of $D0_{19}$ structure and the $L1_2/D0_{19}$ bilayers, respectively. **d,** The calculated interface resistivity in three-dimension $\rho_{xx}^{int}$ (Ωcm) with assumed thickness of interface conducting region $t_{cal}^{int}$. The $\sigma_{sheet}^{int} = \sigma_{sheet_bilayer} - \sigma_{sheet_bulk}$, where $\sigma_{sheet_bilayer}$ and $\sigma_{sheet_bulk}$ are calculated as the average values of two bilayers and two single-layer films ($D0_{19}$ and $L1_2$) shown in **a**, respectively.

**Fig. 4 | Intrinsic-mechanism driven large anomalous Hall conductivity in the kagome-bilayers.** **a, b, and c,** Magnetization $M$ (top) and thickness-averaged Hall conductivity $\sigma_{xy}^{ave}$ (bottom) at $T$ = 300 K of the bilayer, the $D0_{19}$-type film, and the $L1_2$-type film, respectively. The $M$ of the $D0_{19}$-type and $L1_2$-type film is identical to that of ref. 36. **d,** Temperature dependence of $\sigma_{xy}^{ave}$ of the bilayer, the $D0_{19}$-type film, and the $L1_2$-type film. **e,** A scaling plot between $\sigma_{xy}^{ave}$ and $\sigma_{xx}^{ave}$ at $T$ = 300 K for the bilayer, the $D0_{19}$-type film, and the $L1_2$-type film with the values of previously studied materials[26,41]. **f,** Electronic band structures of $D0_{19}$-type $Co_3Mo$, which is obtained from ab initio calculations (left). $E_F$ corresponds to Energy = 0 eV. In **f,** the green, red, and blue open circles represent orbital weights projected onto Co $d_{x^2-y^2} + d_{xy}$, $d_{z^2}$, and $d_{xz} + d_{yz}$ orbitals, respectively. The corresponding energy dependence of the anomalous Hall conductivity for $D0_{19}$-type $Co_3Mo$ (right). Red-shaded region around 0.35 eV corresponds to large $\sigma_{xy}$ of about 1000 $\Omega^{-1}cm^{-1}$. **g**, Expanded band structures corresponding to red square in **f**. Red and blue arrows indicate the gap opening points. **h**, Berry curvature distribution in $k_x$-$k_y$ plane at $E$ = 0.347 eV. Color bar shows degree of Berry curvature.

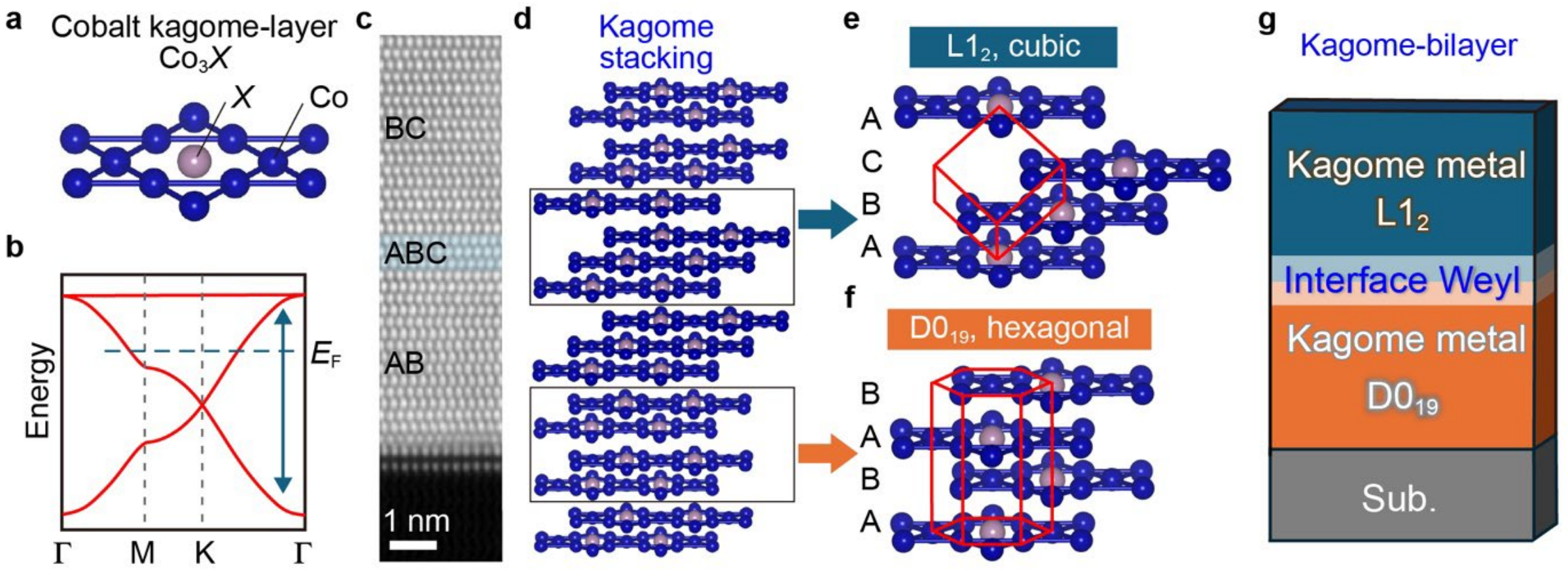

Figure 1 K. Ishida *et al*.,

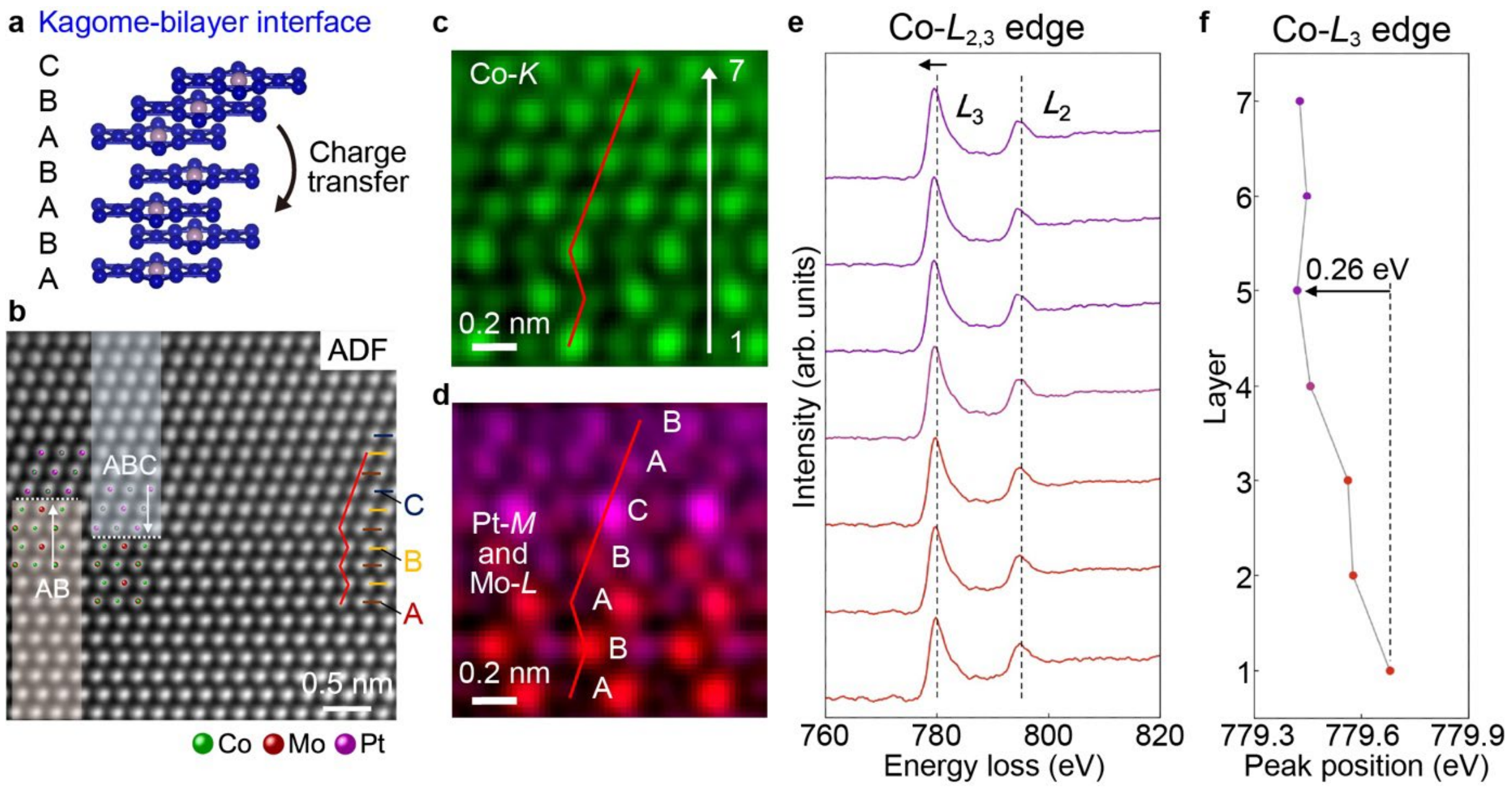

Figure 2 K. Ishida *et al*.,

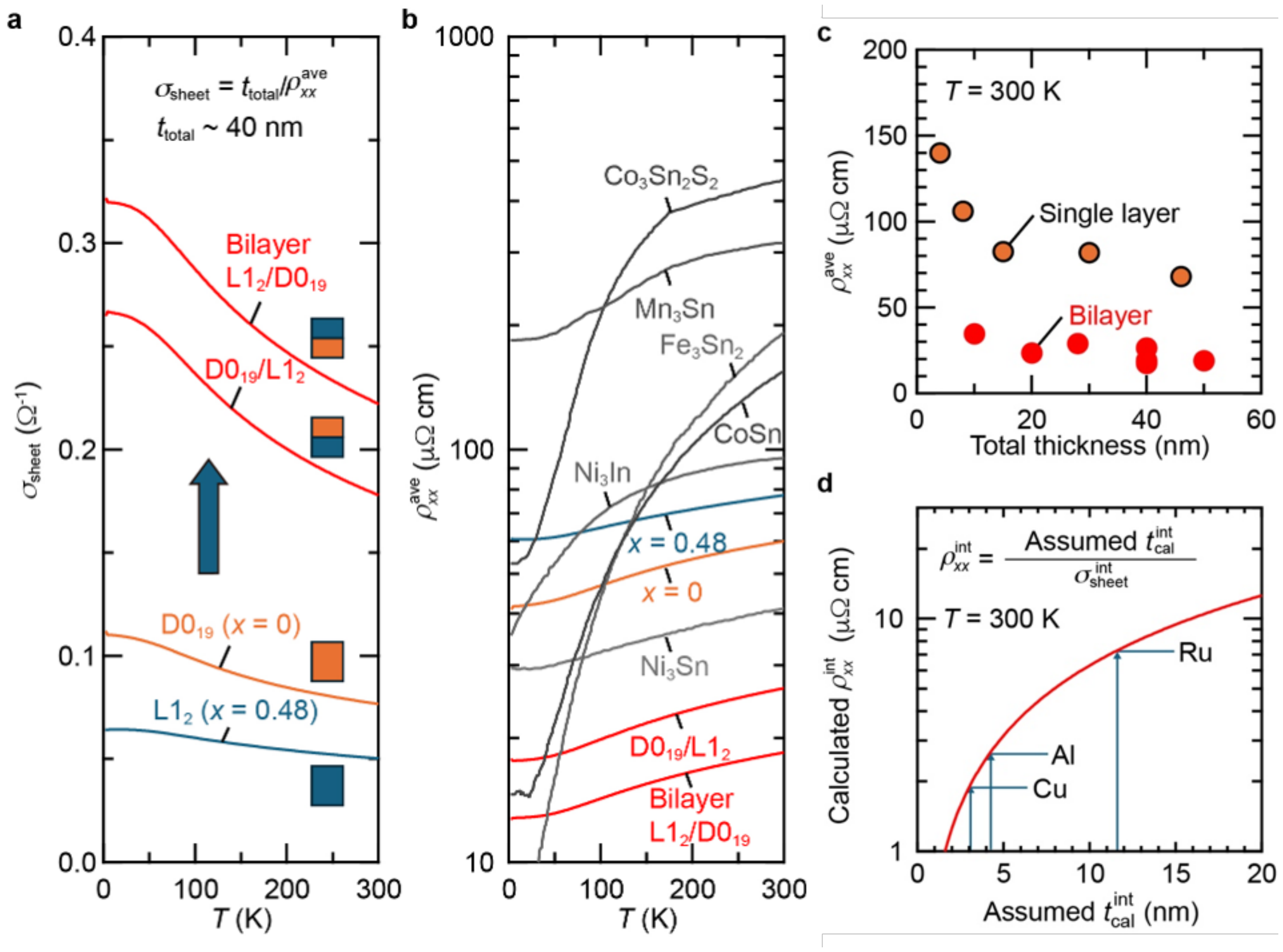

Figure 3 K. Ishida *et al*.,

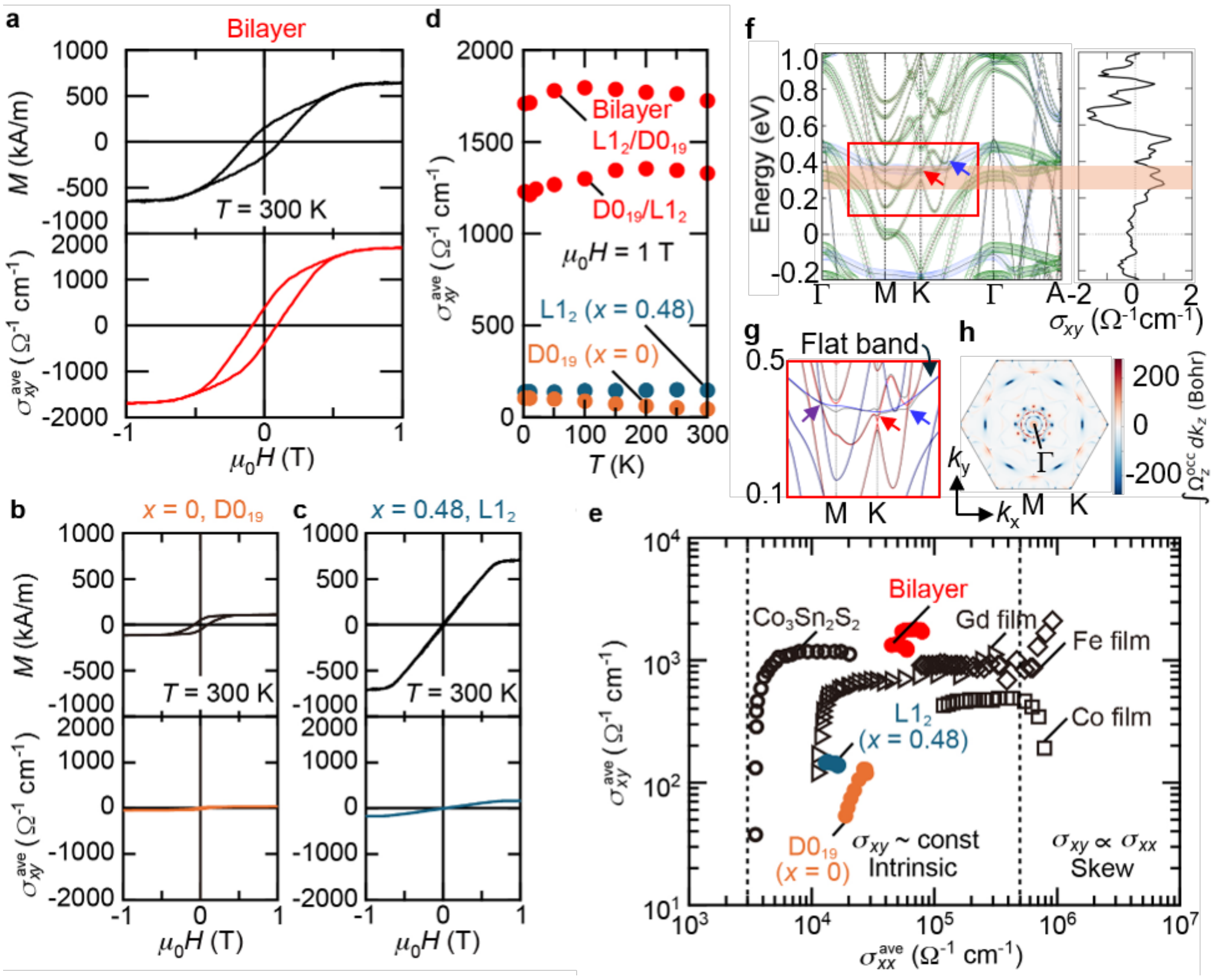

Figure 4 K. Ishida *et al*.,

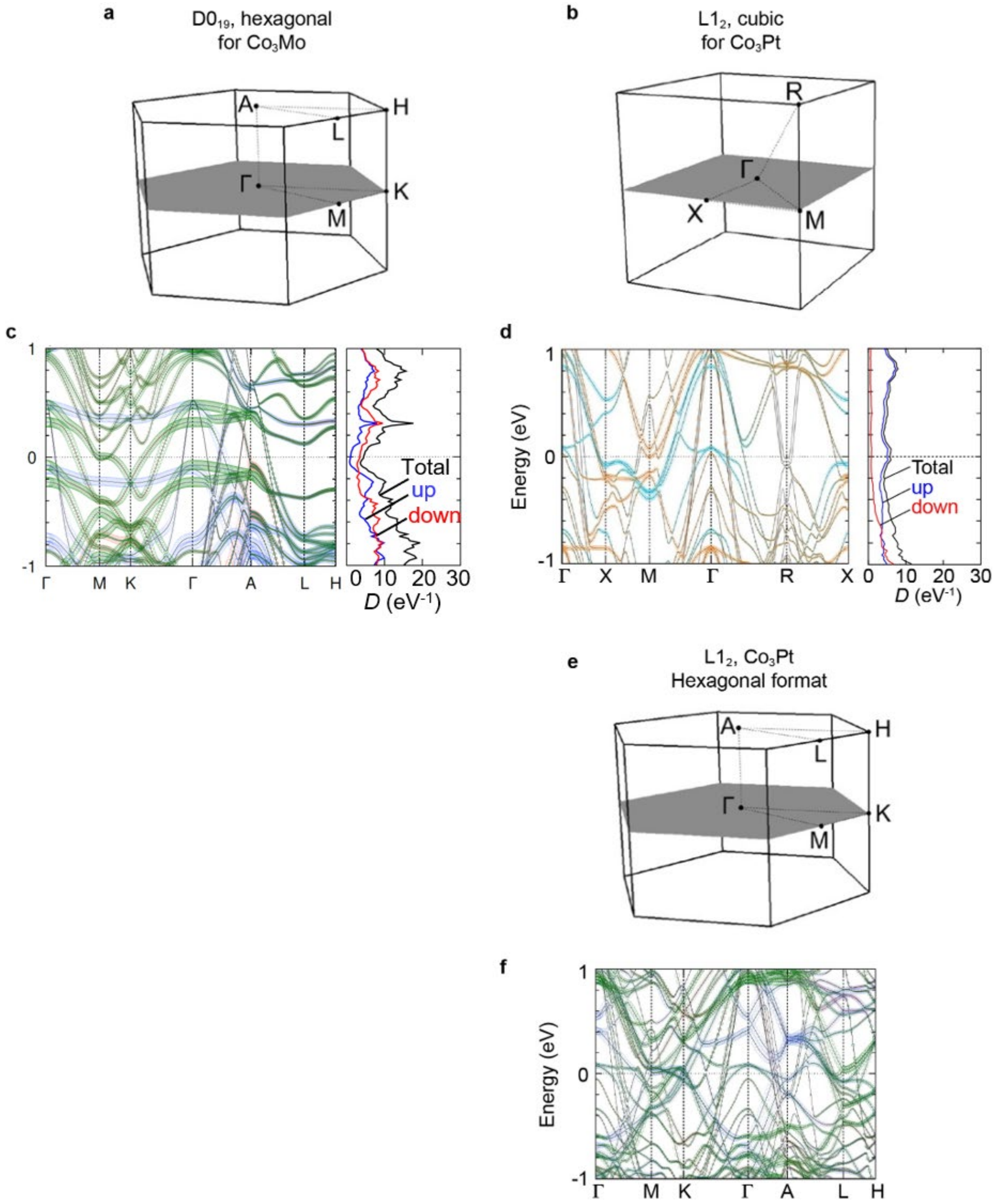

**Extended Data Fig. 1 | a and b,** Schematics of Brillouin zone for $D0_{19}$-type and $L1_2$-type structures, respectively. **c and d,** Electronic band structure (left) and density of states (right) for $D0_{19}$-type (**c**) and $L1_2$-type (**d**) structures. The red, blue, and black lines in density of states correspond to the up spin, down spin and total values, respectively. **e**, Schematics of Brillouin zone depicted by hexagonal symmetry of $Co_3Pt$ of $L1_2$-type structure in **f**.

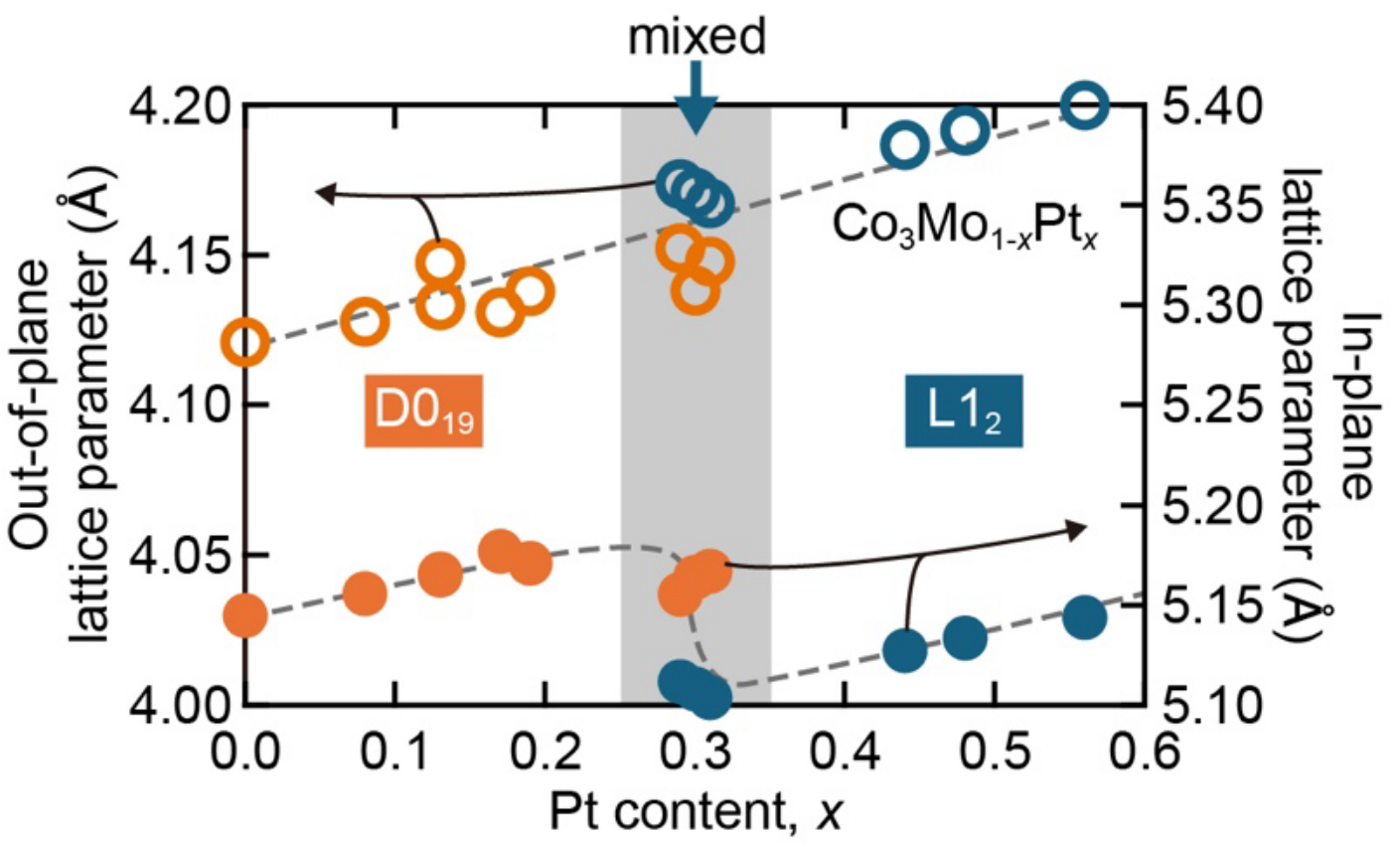

**Extended Fig. 2 |** Out-of-plane (left) and in-plane (right) lattice parameters of $Co_3Mo_{1-x}Pt_x$ films. Structural phase transition from $D0_{19}$ to $L1_2$ in $Co_3Mo_{1-x}Pt_x$ films across mixed-alloy film (gray) region, where both $D0_{19}$ and $L1_2$ diffraction peaks were observed in x-ray diffraction.

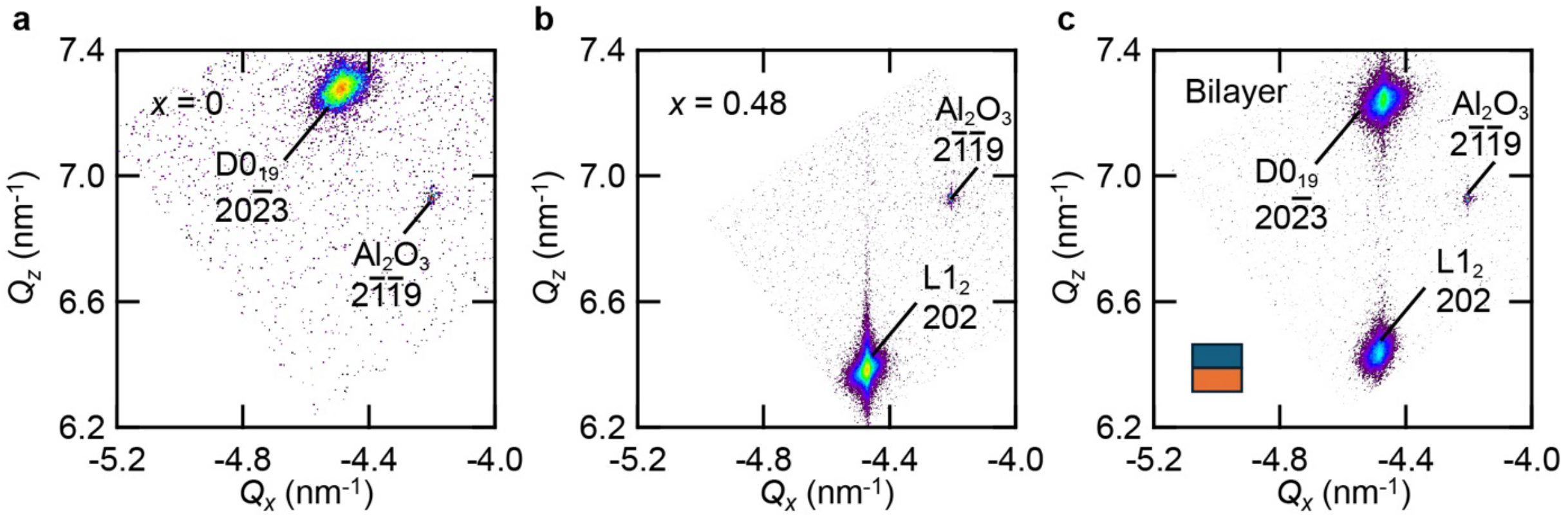

**Extended Data Fig. 3 | a, b, and c,** Reciprocal space mapping of the three samples, $D0_{19}$-type ($x$ = 0, left), $L1_2$-type ($x$ = 0.48, center), and the $L1_2/D0_{19}$ bilayer (right). The in-plane-lattice constant of top $L1_2$ and bottom $D0_{19}$ layer in the bilayer is well matched. These results were measured for the identical samples in Fig. 3a.

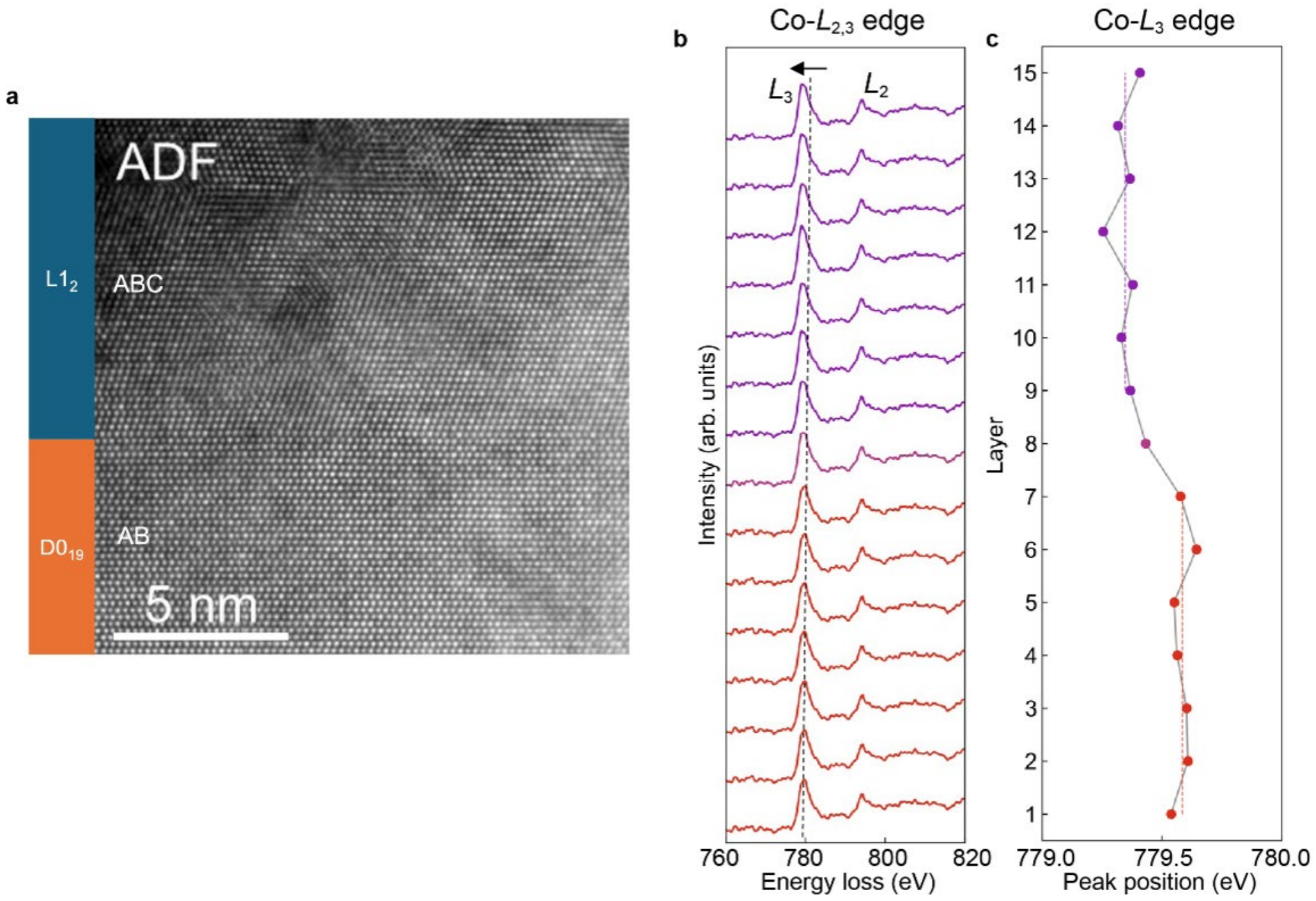

**Extended Data Fig. 4 | a,** Large field-of-view ADF STEM image of the bilayer sample. **b**, EEL spectra of Co-$L_{2,3}$ edges at fifteen points across the interface. **c**, Peak positions of Co-$L_3$ edges across the interface.

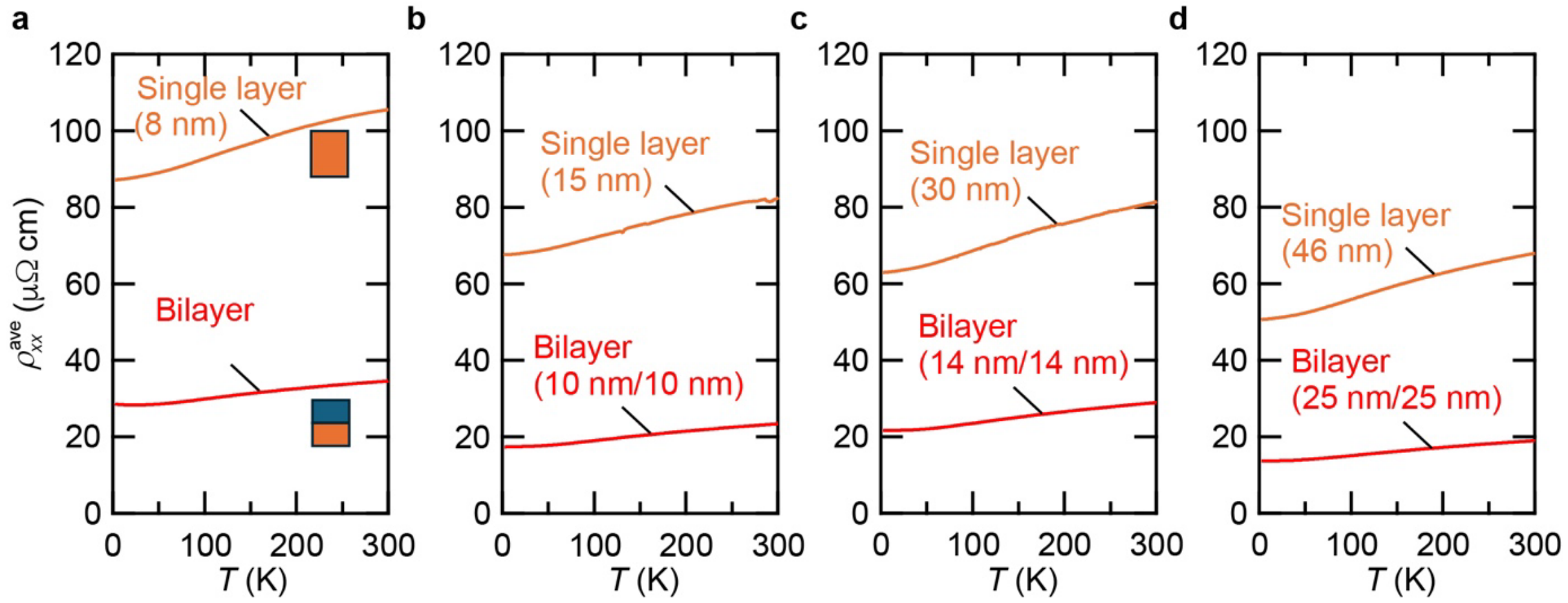

**Extended Data Fig. 5 | a, b, c, and d,** Temperature dependence of the average resistivity for the single layers of $D0_{19}$ ($t$ = 8, 15, 30, 46 nm) and $L1_2/D0_{19}$ bilayers ($t_{total}$ = 10, 20, 28, 50 nm).

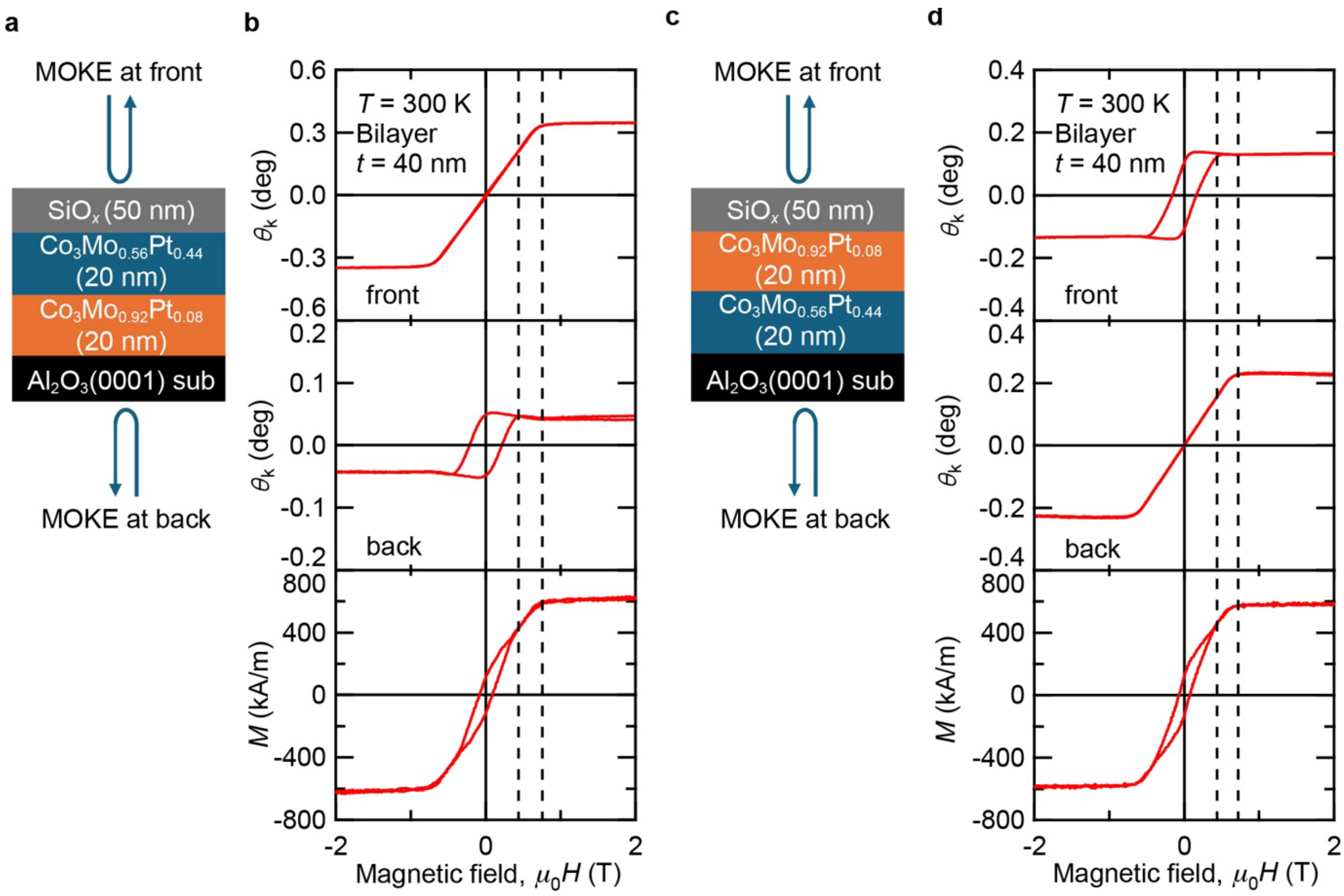

**Extended Data Fig. 6 | a and c,** Schematics of measurement configuration for magneto-optical Kerr effect (MOKE). Stacking sequence of kagome-bilayers is reversed, **a**, $L1_2/D0_{19}$ and **c**, $D0_{19}/L1_2$. **b and d,** Top and middle panels show polar MOKE curves at $T$ = 300 K measured by the laser incident from the frontside- and backside of the bilayer, respectively, as a function of out-of-plane magnetic field. Upper (Lower) $L1_2$-type layer has in-plane magnetization direction. Lower (upper) $D0_{19}$-type layer has perpendicular magnetization direction. These magnetization processes are consistent to that observed for the single-layer films. Bottom panel shows magnetization curve of the bilayer. The broken lines denote the saturation fields for the $L1_2$-type layer and $D0_{19}$-type layer.

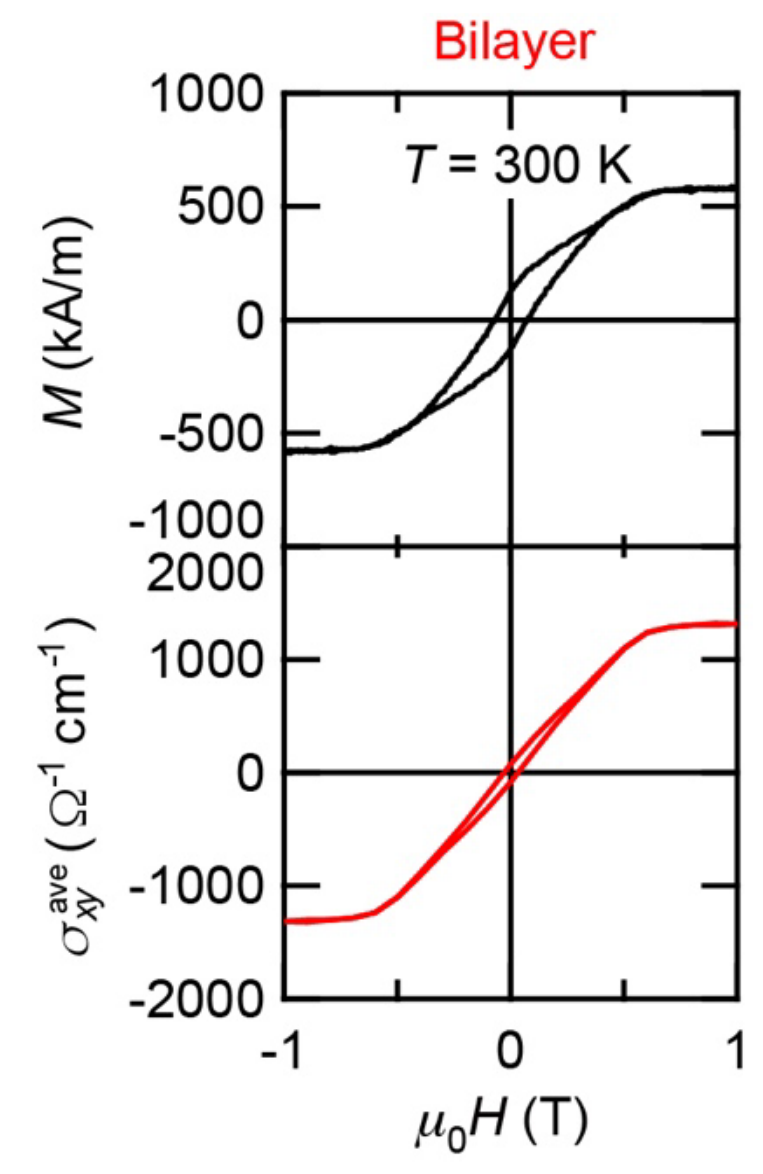

**Extended Data Fig. 7 |** Magnetization $M$ (top) and thickness-averaged Hall conductivity $\sigma_{xy}^{\mathrm{ave}}$ (bottom) at $T$ = 300 K of the $D0_{19}/L1_2$ bilayer.

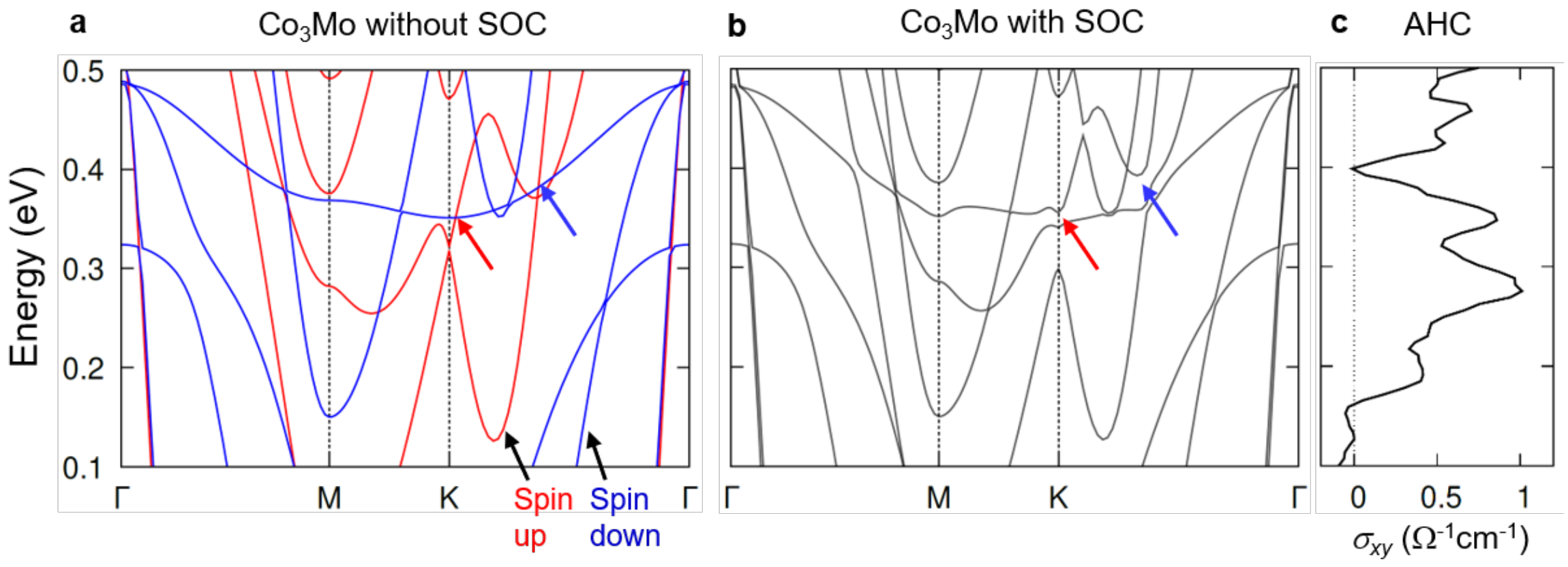

**Extended Data Fig. 8 | a and b,** Calculated band structure of $Co_3Mo$ with and without spin-orbit coupling (SOC), respectively. **c**, Calculated $\sigma_{xy}$ corresponding energy region.

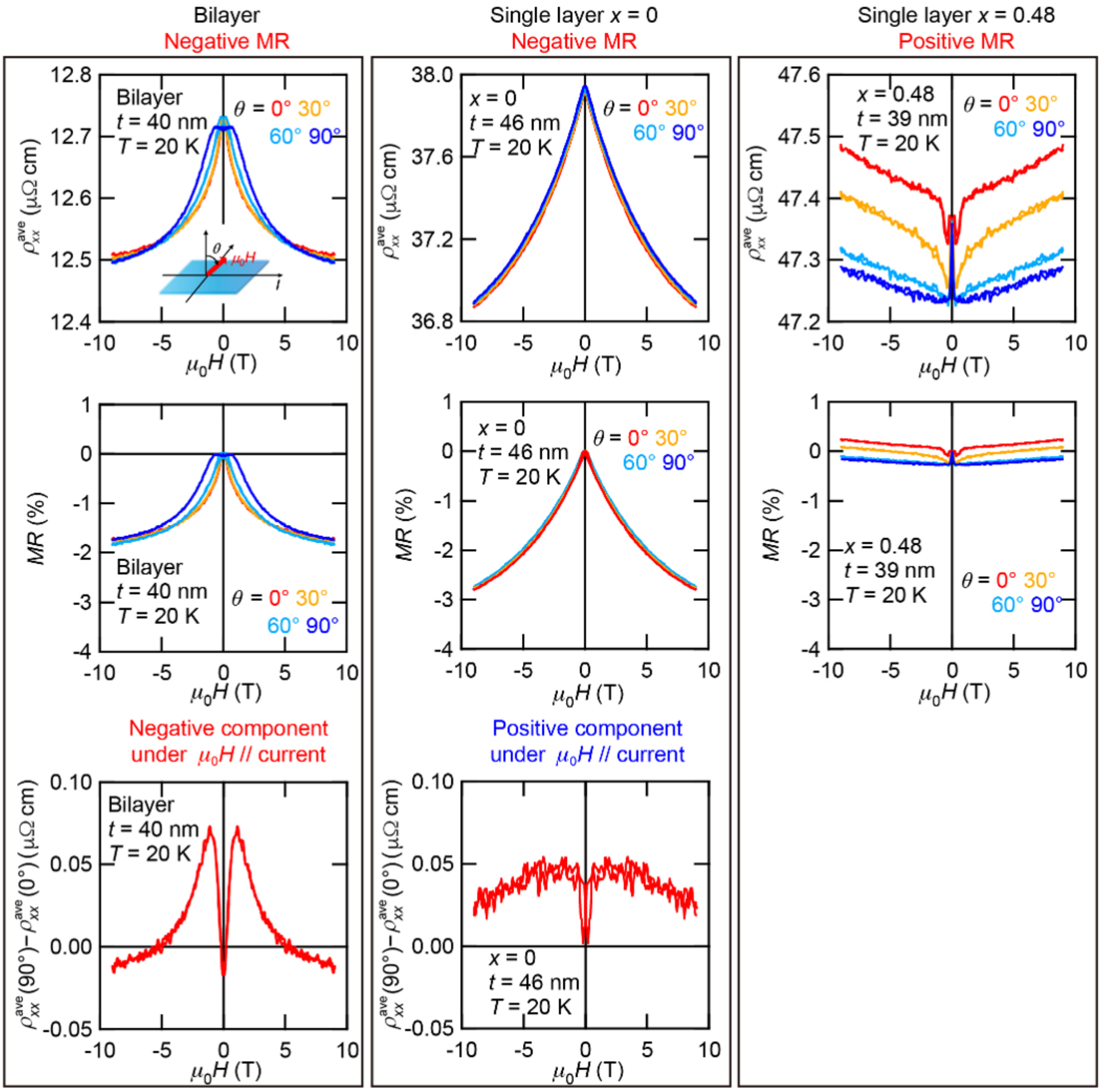

**Extended Data Fig. 9 | a-c,** Magnetoresistance plots for (left) the $L1_2/D0_{19}$ bilayer, (center) the single layer of $D0_{19}$ with $x = 0$, and (right) the single layer of $L1_2$ with x = 0.48. ((Top) Raw data of magnetoresistance under various magnetic field direction at $\theta = 0^o$, $30^o$, $60^o$, $90^o$ at $T =$ 20 K. (Center) Magnetoresistance *MR* calculated by $MR(\%) = \frac{\rho_{xx}^{ave}(\mu_0 H) - \rho_{xx}^{ave}(\mu_0 H = 0\ T)}{\rho_{xx}^{ave}(\mu_0 H = 0\ T)} \times 100$. (Bottom) The difference of magnetoresistance between parallel ($90^o$) and perpendicular ($0^o$). Negative value at high field in the bilayer likely links to the chiral anomaly.